\documentclass[11pt]{article}

\usepackage{cite}
\usepackage{amsfonts}

\usepackage[pdftex]{graphicx}
\usepackage{epstopdf}
\usepackage{amsmath}
\usepackage{amssymb}
\usepackage{tikz}
\usepackage{lineno}

\usepackage[final]{pdfpages}

\newcommand{\ka}{\kappa}

\newcommand{\taur}{\tau_{\rm relax}}
\newcommand{\taunew}{t_f}

\definecolor{linkcolor}{rgb}{0,0,0.6} 
\usepackage[pdftex,colorlinks=true, pdfstartview=FitV, linkcolor= linkcolor, citecolor= linkcolor, urlcolor= linkcolor, hyperindex=true,hyperfigures=true]{hyperref} 

\begin{document}

\title{
  {\bf Engineered Swift Equilibration of a Brownian particle} }

\author{Ignacio A. Mart\'inez $^1$, Artyom Petrosyan $^1$, 
David Gu\'ery-Odelin$^2$, \\  Emmanuel Trizac$^3$, Sergio Ciliberto$^1$\\
\\  $      \  \  $
$1:$ Laboratoire de Physique, CNRS UMR5672 \\  Universit\'e de Lyon, \'Ecole Normale Sup\'erieure,  \\
46 All\'ee d'Italie, 69364 Lyon, France. \\
$2:$ Laboratoire Collisions Agr\'egats R\'eactivit\'e,   CNRS UMR5589,\\ Universit\'e de Toulouse, 31062 Toulouse, France  \\ $3:$ LPTMS,  CNRS,  Univ. Paris-Sud,  Universit\'e Paris-Saclay, \\ 91405 Orsay,  France }

\date{\today}

\maketitle

\textbf{  {A fundamental and intrinsic property of any device or natural system is its relaxation time τrelax, which is the time it takes to return to equilibrium after the sudden change of a control parameter \cite{boltzmann1964}. Reducing $\taur$, is frequently necessary, and is often obtained by a complex feedback process. To overcome the limitations of such an approach, alternative methods based on driving have been recently demonstrated \cite{torrontegui2013,PRX2014}, for isolated quantum and classical systems \cite{CKR08,SSV10,SSC11,BVM12,Bowler,Walther}. Their extension to open systems in contact with a thermostat is a stumbling block for applications. Here, we design a protocol,named Engineered Swift Equilibration (ESE), that shortcuts time-consuming relaxations, and we apply it to a Brownian particle trapped in an optical potential whose properties can be controlled in time. We implement the process experimentally, showing that it allows the system to reach equilibrium times faster than the natural equilibration rate. We also estimate the increase of the dissipated energy needed to get such a time reduction. The method paves the way for applications in micro and nano devices, where the reduction of operation time represents as substantial a challenge as miniaturization \cite{peercy2000}.
}}

The concepts of equilibrium and of transformations from an equilibrium state to another, are  cornerstones of thermodynamics. A textbook illustration  
is provided by the expansion of a gas, starting at equilibrium and expanding to reach a new equilibrium in a larger vessel. This operation can be performed  
either very slowly by a piston,  without dissipating energy into the environment, or alternatively quickly, letting the piston freely move to reach the new volume. 
In the first case, the 
transformation takes a long (virtually infinite) time to be completed, while the gas is always in a quasi-equilibrium state.  
In the second case instead, the transformation  
is fast but the gas takes its characteristic relaxation time $\taur$  to reach the new equilibrium state in the larger volume.  
This is the time required for the exploration of the new vessel. 
More generally, once a control parameter is suddenly changed, the accessible phase space changes too \cite{boltzmann1964,maxwell1867}; 
the system adjusts and needs a finite time to reach the final equilibrium distribution.
This equilibration process  plays of course a key role  in out of equilibrium 
thermodynamics.

An important and relevant question related to optimization theory is whether  
a targeted  statistical equilibrium state can be reached in a chosen time, arbitrarily shorter than $\taur$. 
Such strategies are reminiscent of those worked out in the recent field of Shortcut to Adiabaticity \cite{torrontegui2013,PRX2014};  they aim at developing	 protocols,  
both in  quantum and in  classical regimes,  allowing the system to move as fast  as possible  from one equilibrium position to a new one, 
provided that there exist an adiabatic transformation relating the two \cite{chen2010,GMR14,PaSt15}. 
So far, proof of principle experiments have been carried out for isolated systems \cite{CKR08,SSV10,SSC11,BVM12,Bowler,Walther} 
and for photonics circuit design \cite{TsC12,Tse14,HTs14,Ste14}.
Yet, the problem of open classical systems is untouched. We solve here this question by putting forward an accelerated 
equilibration protocol for a system in contact with a thermal  bath.  {Such a protocol shortcuts quasi-stationarity, 
according to which a driven open system remains in equilibrium with its environment at all times.}
This is a key step for a number of applications in nano oscillators \cite{kaka2005}, 
in  the design of nanothermal engines \cite{martinez2015}, or in monitoring mesoscopic chemical or
biological processes \cite{collin2005}, for which thermal fluctuations are paramount and 
an accelerated equilibration desirable for improved {power}.
We dub the method Engineered Swift Equilibration (ESE).

However, an arbitrary reduction of the time to reach equilibrium will have unavoidable consequences from an energetical point of view \cite{cui2015}.
The question of the corresponding cost is relevant as such, but also for applications, for example  in nano-devices \cite{peercy2000} where the 
goal is the  size and execution time reduction of a given process. 
Here, beyond the theoretical derivation of the procedure,
we develop an experimental demonstration of ESE, studying the dynamics of a colloidal particle within an optical potential. 
The energetics of the system will also be analyzed in depth, shedding  light on the inherent consequences of time-scale reduction \cite{cui2015,ScSe07,ScSe08,Aurell,Deff15,Pol15}.

\begin{figure}[!ht]
\centering
\includegraphics[width=0.8 \columnwidth]{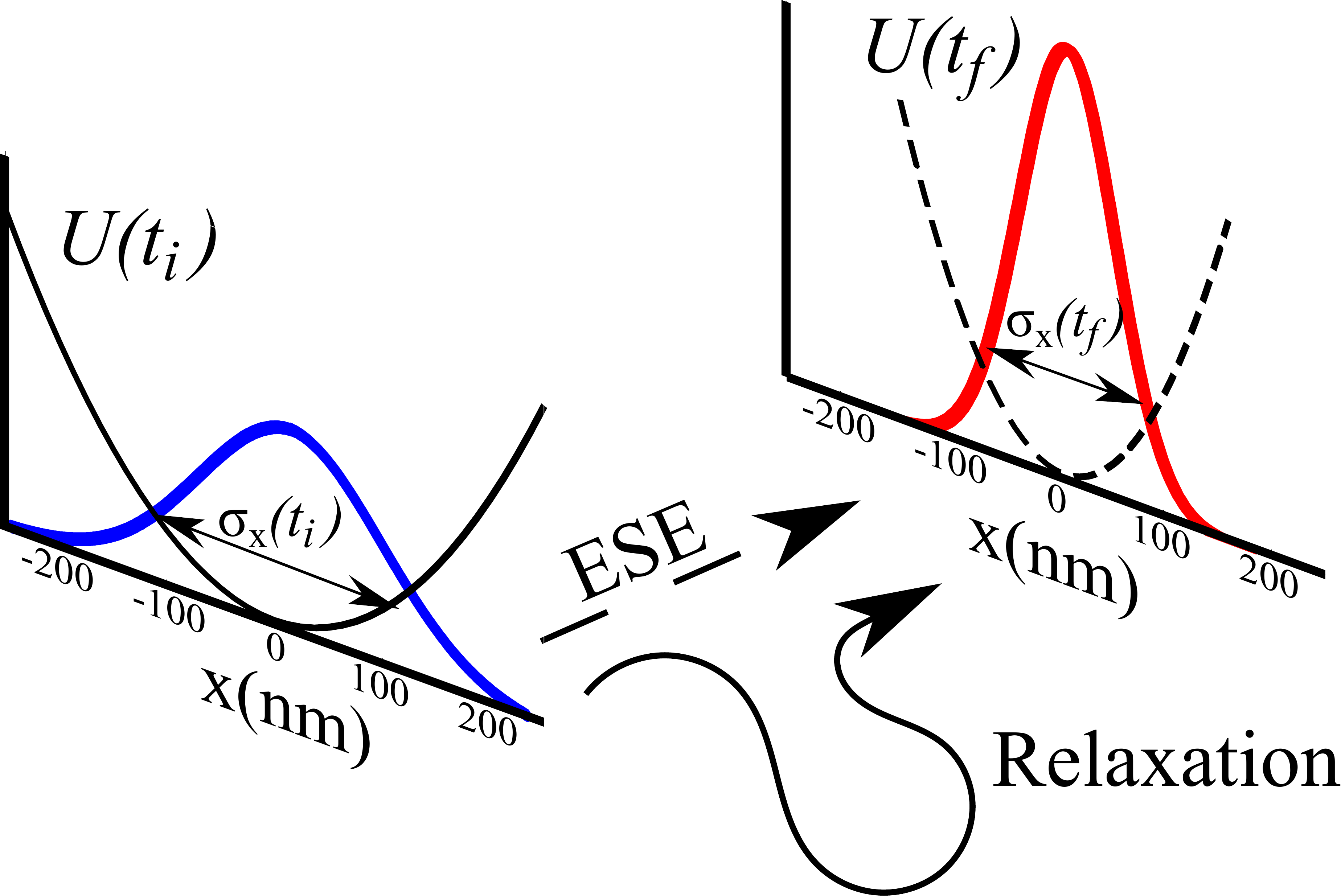}
\caption{\textbf{Sketch of the  process.}
At initial time $t_i$, the particle is at equilibrium, confined in a potential 
of stiffness $\ka_i$ (black line), and $\rho(x)$ (blue histogram) has variance $\sigma_x^2(t_i)=k_BT/\kappa_i$.  
After a long relaxation where $\kappa$ is gradually increased, 
the particle is at time $t_f$ at equilibrium in  a 
stiffer potential (black line). Since  $\ka_f>\ka_i$, the variance $\sigma_x^2(t_f)$ of position (red histogram) 
is smaller than its initial counterpart. 
The goal is to work out a protocol with a suitable dynamics $\ka(t)$, that would ensure equilibrium at an arbitrary
chosen final time $t_f$, no matter how small. }
\label{fig:intro}
\end{figure}

Our experimental system consists of a microsphere immersed in water \cite{neuman2004} (see  Methods \ref{Methods}). The particle is trapped by an optical harmonic potential $U(x,t)=\kappa(t)x^2/2$, where $x$ is the particle position and $\kappa(t)$ is the stiffness of the potential which can be controlled by the power of the trapping laser \cite{martinez2015}. The system is affected by thermal fluctuations; its  dynamics is overdamped and described by a Langevin equation. 
Our Brownian particle has a relaxation time defined as $\taur=\gamma/\kappa$, where  
$\gamma$ is the fluid viscous coefficient.
At equilibrium, the probability density function (pdf) $\rho(x)$ of $x$  is Gaussian 
$\rho_{\rm eq}(x) = 1/\sqrt{\pi \sigma_x^2}\exp(- x^2/(2\sigma_x^2)$ with variance $\sigma_x^2= k_BT/\ka$ as prescribed by the equipartition theorem. Here $k_B$ is the Boltzmann constant and $T$ the bath temperature. 
In this system, we consider  the compression process sketched in Fig. \ref{fig:intro}, in which  the stiffness is changed from an initial value to a larger one. The evolution of the system during the relaxation towards the new equilibrium state  
is monitored through the position pdf $\rho(x,t)$, which is Gaussian at all times {(see Methods and Supplementary Information)}. Thus, the distribution $\rho(x,t)$ is fully characterized by the time evolution of its mean and its standard deviation $\sigma_x (t)$. The main question is now that of finding, provided it exists, a suitable time evolution of  the stiffness $\kappa(t)$ (our control parameter),
for which the equilibration process  is much faster than $\taur$. This question can be affirmatively answered using a particular solution of the Fokker-Planck equation (see Methods \ref{sec_ESE} 
and  Supplementary Information).
{We emphasize that the ESE idea is not restricted to manipulating Gaussian states and that non harmonic 
potentials $U$ can be considered, along the lines presented in the Supplementary Information}.

\begin{figure}[!ht]
\centering
\includegraphics[width=0.8\columnwidth]{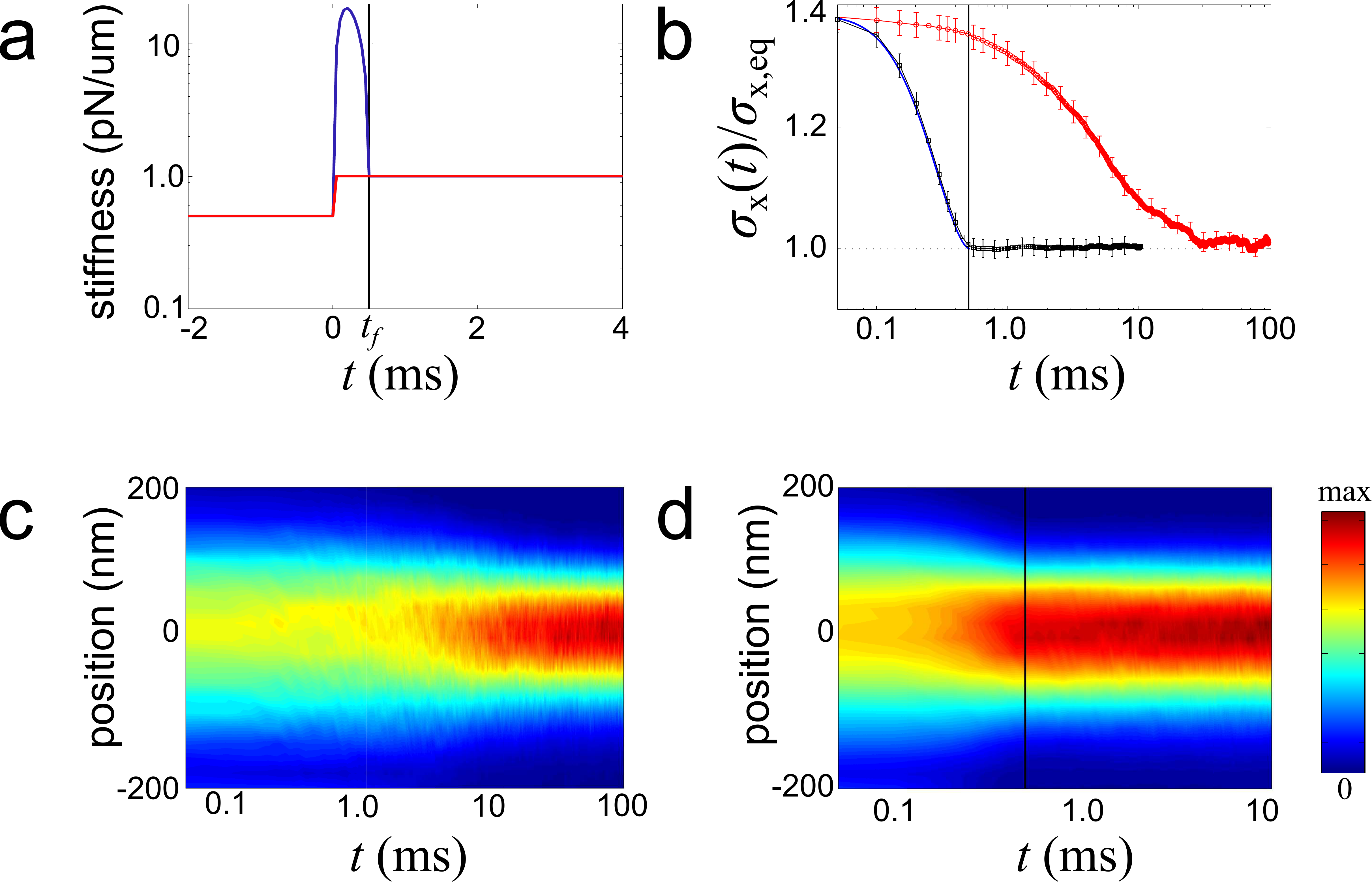} 
\caption{\textbf{Dynamics of the system along the STEP and ESE protocol.} 
\textbf{a}- Experimental protocols: STEP route (red) and ESE route (blue). 
The system starts with $\kappa_i =0.5$ pN/$\mu$m at $t=0$
to finish with $\kappa_f =1.0$ pN/$\mu$m. In all  figures, the vertical solid line at $t=\taunew$ indicates the end of 
the ESE protocol.
\textbf{b}- Normalized standard deviation $\sigma_x (t)$ of the particle's trajectory 
along the STEP (red circles) and ESE protocol (black squares). 
The blue solid line represents the theoretical prediction of the variance evolution,
{i.e. $1/(2\alpha)$ where $\alpha$ is given by Eq. \eqref{eq:alpha}.  {The error bars take into account the calibration and the statistical errors.} }
\textbf{c}- Time evolution of the position pdf.  
The color map of $\rho(x,t)$ is plotted after an instantaneous change of the stiffness at $t=0$ (STEP).  
\textbf{d}- ESE counterpart of panel c. }
\label{fig:colormap}
\end{figure}

In this Letter, two methods are compared. On the one hand, at a given instant $t_i=0$, we suddenly change $\kappa$ from the initial value $\ka_i$ to the final value  $\ka_f$. During this protocol, referred to as STEP, the particle mean position 
does not change while the spread $\sigma_x$ equilibrates to the new value $\sqrt{k_BT/\ka_f}$ in about  3  relaxation 
times $\taur= \gamma/\ka_f $. On the other hand, following the ESE procedure, $\ka(t)$ is modulated in such a way that $\sigma_x$  is fully equilibrated  at 
$\taunew \ll \taur$. The protocol which meets our requirements is 
given by eq. \eqref{eq:protocole} (in Methods).
In the experiment, we select $\kappa_i=0.5$ pN/$\mu$m and 
$\kappa_f=1.0$ pN/$\mu$m  in such a  way that $\taur \simeq 15$ ms. Furthermore, in order to have  a well defined separation between time scales, we choose $\taunew= 0.5$ ms, which 
is roughly 100 times smaller than the thermalization time in the STEP protocol.  Both protocols are displayed in Fig. \ref{fig:colormap}a where we can appreciate the rather complex time dependence of the ESE control procedure.
This is a necessity to allow for a quick evolution to the new equilibrium state. The faster the evolution (smaller $t_f$), the stiffer the transient confinement must 
be (the maximum stiffness reached in Fig. \ref{fig:colormap}a  is $37 \,\kappa_i$). 
In order to study the evolution of $\rho(x,t)$ for  the two protocols, we perform the following cycle. 
First, the particle is kept at $\kappa_i$ for 50 ms to ensure
proper equilibration. Then, at $t=0$ ms we apply the protocol (either STEP or ESE) and $x(t)$ is measured for $10$ ms in the case of ESE and 100 ms for STEP. Finally, the stiffness 
is set again to $\ka_i$ and this  cycle is repeated $N$ times. 
The evolution of $\sigma_x (t)$ for $t>0$  is obtained by performing an ensemble average over
$N=$2 $10^4$ cycles.

The results are shown in Fig. \ref{fig:colormap}b, where $\sigma_x(t)$ is plotted as a function of time  for the two protocols, 
from one equilibrium configuration to the other. It appears that the engineered system reaches the target
spread precisely at $t_f$ and subsequently does not evolve.
On the other hand, the STEP equilibration occurs after a time close to  $3 \,\tau_{\rm relax}$. 
Figures \ref{fig:colormap} c-d represent the complete STEP and ESE dynamics of $\rho(x)$. 
The Gaussian feature is confirmed experimentally during  ESE, even far from equilibrium, 
since the kurtosis is ${\rm Kurt} (x)=(3.00\pm0.01)$.
The results of Fig. \ref{fig:colormap} clearly show the efficiency of ESE, driving the system into equilibrium in a time which is 100 
times shorter than the nominal equilibration time $3 \,\taur$.

\begin{figure}[!h]
\centering
\includegraphics[width=0.8\columnwidth]{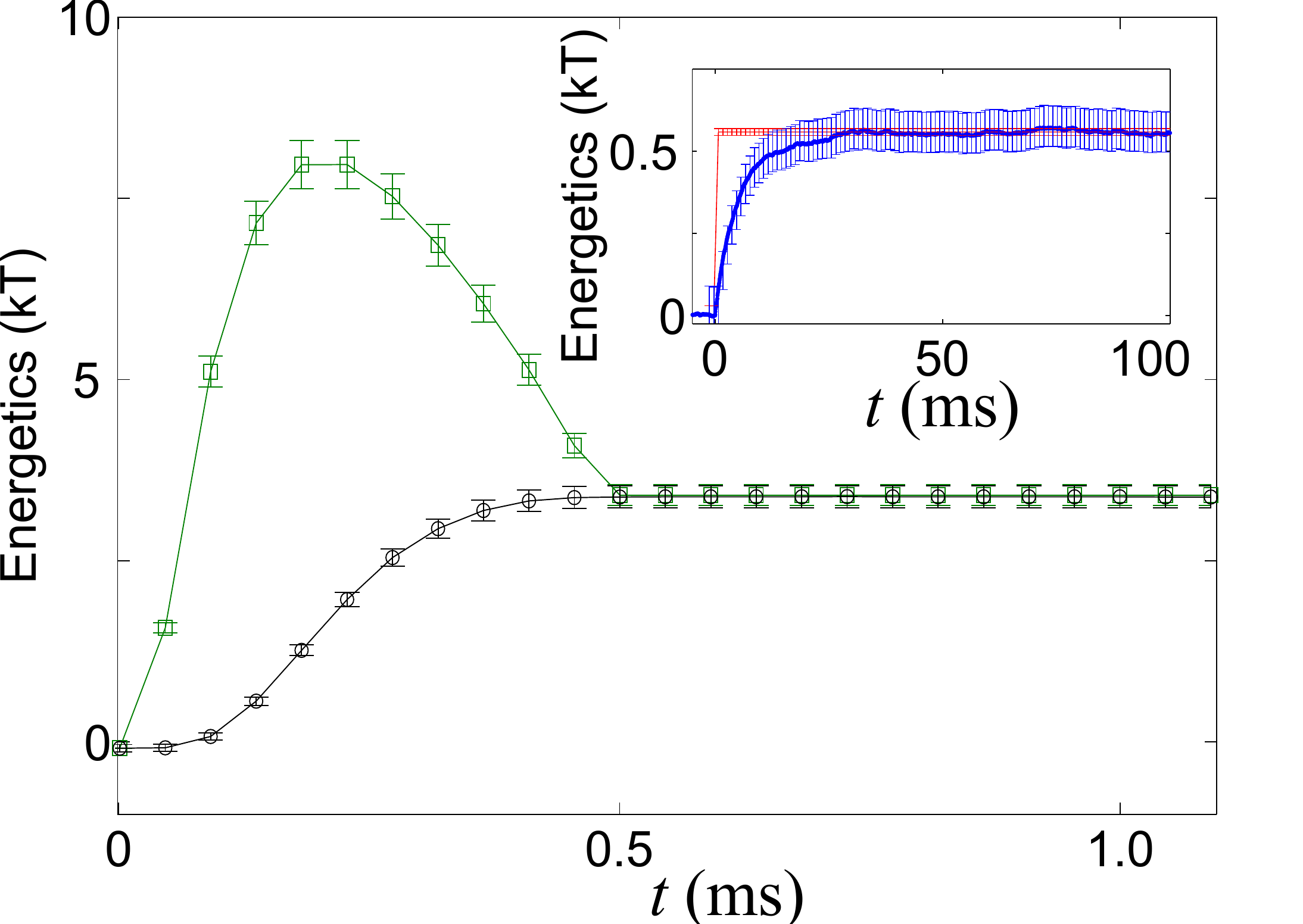} 
\caption{\textbf{Energetics of the ESE protocol.} Average value of the cumulative work $W$ (green squares) and heat $Q$ (black circles) are represented as a function of time. 
The energy exchange stops within the protocol time $\taunew=0.5\,$ms. For $t>\taunew$, $\langle W\rangle = \langle Q\rangle$: the system is in an equilibrium steady state, in contact with an  
isothermal reservoir. \textbf{Inset}. Energetics of the STEP protocol.  Work (red curve) is exerted onto the system quasi-instantaneously, with an abrupt change of trap stiffness. 
On the other hand, heat (blue curve) is delivered along the whole equilibration process. 
{  {The error bars, which  take into account the calibration and the statistical errors, }have the same relative value for ESE and STEP }. Energy-wise, the ESE method appears more costly: 
this is the price for accelerating the thermalization process.
}
\label{fig:energetics}
\end{figure}

We now turn our attention to the energy required  for achieving such a large time reduction. 
Developments in the field of stochastic thermodynamics \cite{seki2010} endow work $W$ and heat $Q$ with a clear mesoscopic 
meaning, from which a resolution better than $k_BT$ can be achieved experimentally
(see Methods for an explicit definition). In Fig. \ref{fig:energetics}, the complete 
energetics of our system is shown 
for the ESE and STEP protocols. The evolution of the mean cumulative work $\langle W (t)\rangle $ reveals the physical behavior of the system undergoing ESE. 
In the first part of the protocol ($t<0.2\,$ms), confinement is increased which provides positive work to the system. In the subsequent evolution ($0.2<t<0.5\,$ms),
work is delivered from the system to the 
environment through the decrease of the stiffness. In striking contrast with an adiabatic transformation, the value of heat increases  monotonically, as the system dissipates 
heat all over the protocol. In the inset of Fig. \ref{fig:energetics}, $\langle Q\rangle$ and $\langle W\rangle$ are shown for the STEP process. 
Notice how the work exerted on the system is 
almost instantaneous, while heat is delivered over a wide interval of time, up to complete equilibration. 
Quite expectedly, there is a price to pay for ESE. A significant amount of work is required
to speed up the evolution and beat the natural time scale of our system \cite{cui2015}. It can be shown that the cost $\langle W(t_f)\rangle$
behaves like $\taur/t_f$ for $t_f \to 0$. More precisely, this amounts to a time-energy uncertainty relation: 
$t_f \,\langle W(t_f)\rangle \sim 0.106\, (2\taur) \, k_B T$.
If instead, one proceeds in a quasi-static fashion ($t_f\gg \taur$), 
the cost reduces to the free energy difference, $k_BT \log (\kappa_f/\kappa_i )/2$ which is $0.35 \,k_BT$ when $\kappa_f=2\kappa_i$.
For the ESE experiments shown, we have $\langle W(t_f)\rangle\simeq 3.5\,k_BT$, about 10 times larger.

Our results show the feasibility and expediency of accelerated protocols for equilibrating confined 
Brownian objects. The ESE path allowed to gain two orders of magnitude in the thermalization 
time, as compared to an abrupt change of control parameter (STEP process). The associated 
energetic cost has been assessed.  Finally, while an over-damped problem has been solved here,
{the generalization of the ESE protocol to non-isothermal regimes for under-damped systems can in principle be worked out theoretically. 
Its application to AFM tip,  
vacuum optical traps, or to transitions between non-equilibrium steady states, constitutes a 
timely experimental challenge in this emerging field.} 

  {{\bf Author contributions:} All authors contributed substantially to this work.}

{\bf Acknowledgements.} 
{We would like to thank B. Derrida for useful discussions.} I.A.M., A.P. and S. C. acknowledge financial support from the European Research Council Grant OUTEFLUCOP.

\bibliographystyle{ieeetr}
\bibliography{references_ese}

\section*{Methods}

\subsection{\textit{Experimental setup.}} \label{Methods}
Silica microspheres of radius 1$\mu$m were diluted in milliQ water to a final concentration of a few spheres per milliliter. The microspheres were inserted into a fluid chamber, which 
can be displaced in 3D by a piezoelectric device (Nanomax TS MAX313/M). The trap is realized using  a near infrared laser beam (Lumics, $\lambda$=980\,nm with maximum power 500\,mW) expanded and inserted through an 
oil-immersed objective (Leica, 63$\times$ NA 1.40) into the fluid chamber. The trapping laser power, which determines the trap stiffness, is modulated by a external voltage $V_{\kappa}$ via  a Thorlabs ITC 510 laser diode controller 
with a switching frequency of 200 kHz . $V_{\kappa}$ is generated by a National Instrument card (NI PXIe-6663) managed by a custom made Labview program.  The detection of the particle position is  performed using 
an additional HeNe laser beam ($\lambda$=633nm), which  is expanded and collimated by a telescope and passed through the trapping objective. The forward-scattered detection beam is 
collected by a condensor (Leica, NA 0.53), and its back focal-plane field distribution projected onto a custom Position Sensitive  Detector (PSD from First Sensor 
with a  band pass of 257 kHz) whose signal is acquired at a sampling  rate of 20 kHz with a NI PXIe-4492 acquisition board.

\subsection{Energetics measurement.}
From the experimental observables, the stiffness $\kappa$ and the particle position $x$, it is possible to infer the energetic evolution of our system  within the stochastic energetics framework \cite{seki2010}. The notion of work $W$ is related to the energy exchange stemming from the modification of a given control parameter, here the trap stiffness.  
Alternatively, heat $Q$ pertains 
to the energy exchanged with the environment, either by  dissipation or by Brownian fluctuations. The work $W(t)$ and dissipated heat $Q(t)$  are expressed as $W(t)=\int_0^t\frac{\partial U}{\partial \kappa}\circ  {{\rm d}\kappa \over {\rm d}t'}\rm{d} t'$, $Q(t)=-\int_0^t  \frac{\partial U}{\partial x}\circ {{\rm d}x \over {\rm d}t'}\rm{d} t'$ 
where $\circ$ denotes Stratonovich integral and $U$ is the potential energy. Under this definition, the first law reads as $\Delta U(t)=W(t)-Q(t)$, where $W(t), Q(t)$ and $\Delta U(t)$ are 
fluctuating quantities. Since $T$ is fixed, both ESE and STEP processes share the same value 
$\langle\Delta U(t_f)\rangle =0$ between the initial and the final state. As a consequence, 
we have $\langle W(t_f)\rangle=\langle Q(t_f)\rangle$.

\subsection{\textit{ESE protocol for the harmonic potential}} \label{sec_ESE}

{Although the idea is general (as discussed in Supplementary Information), we first 
start by a presentation applying the method to harmonic confinement.}
The dynamics of the system is then ruled by the Langevin equation 
\begin{equation}
\dot{x} \,=\, -\frac{\kappa(t)}{\gamma}x+\sqrt{D} \xi(t)
\label{eq:langevineq}
\end{equation}
where a dot denotes time derivative and $x$ is for the position of the Brownian particle.
The friction coefficient $\gamma=6\pi\eta R$ is here constant, $\eta$ being the the kinetic viscosity coefficient and $R$ the radius of the bead. 
The diffusion constant then reads as $D=k_BT/\gamma$. 
The stiffness $\kappa$ has an explicit dependence on time and $\xi(t)$ is a white Gaussian noise with autocorrelation $\langle \xi(t) \xi(t+t')\rangle=2\delta(t')$. 
Equation \eqref{eq:langevineq} is over-damped (there is no acceleration term in $\ddot x$),
which is fully justified for colloidal objects \cite{BaHa}.
The Langevin description \eqref{eq:langevineq} can be recast into the following Fokker-Planck equation 
for the probability density\cite{Risken}:
\begin{equation}
\partial_t\rho(x,t) \,= \, \partial_x\left[\frac{\kappa}{\gamma}x\rho\right] \,+\, D\, \partial^2_{xx} \, \rho
\label{eq:FP}
\end{equation}

At initial and final times ($t_i$ and $t_f$), $\rho(x,t)$ is Gaussian, as required by equilibrium.
A remarkable feature of the ESE (non-equilibrium) solution is that for intermediate times, 
$\rho(x,t)$ remains Gaussian,
\begin{equation}
\rho(x,t)=\sqrt{\frac{\alpha(t)}{\pi}}\exp\left[- \alpha(t)x^2\right].
\label{eq:intstate}
\end{equation}
We demand that 
\begin{equation}
\alpha(0) = \frac{\kappa_i}{2 k_B T} ~ \hbox{ and } ~
\alpha(t_f) = \frac{\kappa_f}{2 k_B T}.
\end{equation}
Combining eq. \eqref{eq:FP} with eq. \eqref{eq:intstate}, we obtain
\begin{equation}
\left[\frac{\dot{\alpha}}{2\alpha}-\dot{\alpha}x^2\right]\rho=\frac{\kappa}{\gamma}\left(1-2\alpha x^2\right)\rho-2\frac{k_BT}{\gamma}\alpha\left(1-2\alpha x^2\right)\rho.
\label{eq:combFK}
\end{equation}
Requiring that the equality holds for any position $x$, the equation is simplified into:
\begin{equation}
\frac{\dot{\alpha}}{\alpha}=\frac{2\kappa}{\gamma}-\frac{4k_BT\alpha}{\gamma} .
\label{eq:combFKsimply}
\end{equation}
This relation was obtained in \cite{ScSe07,ScSe08} by studying the evolution of the variance $\sigma_x^2$.
However, unlike in these works, we supplement our description with the constraints
$\dot{\alpha}(0)=\dot{\alpha}(t_f)=0$,
as a fingerprint of equilibrium for both $t<0$ and $t>t_f$.

Next, the strategy goes as follows. We choose the time evolution of $\alpha$, complying with the 
above boundary conditions. To this end, a simple polynomial dependence of degree 3 is sufficient.
Other more complicated choices are also possible. 
Introducing the rescaled time $s=t/\taunew$, we have 
\begin{equation}
\alpha(s) \,=\, \frac{1}{2k_BT} \left[ \kappa_i + \Delta \kappa (3 s^2-2 s^3) \right],
\label{eq:alpha}
\end{equation}
where $\Delta\kappa = \kappa_f-\kappa_i$.
Finally, Eq. \eqref{eq:combFKsimply} has been satisfied,
from which we infer the appropriate evolution $\kappa(t)$ that is then implemented in the experiment:

\begin{equation}
 \kappa(t) \,= \, \frac{ 3\gamma \Delta\kappa \, s(1-s)/t_f}{\kappa_i+\Delta\kappa(3s^2-2s^3)}+\kappa_i+\Delta\kappa(3s^2-2s^3).
\label{eq:protocole}
\end{equation}

The analysis, restricted here to the one dimensional problem, can be easily recast in three dimensions.
It is also straightforward to generalize the idea to account for a time-dependent temperature $T(t)$,
which can be realized experimentally \cite{martinez2015}. In this latter situation,
the key relation \eqref{eq:combFKsimply} is unaffected, and therefore indicates
how $\kappa$ should be chosen, for prescribed $\alpha(t)$ and $T(t)$.
This highlights the robustness of the ESE protocol.

{The mean work exchanged in the course of the transformation takes a simple form in our context:
\begin{equation}
\langle W \rangle \,= \, \int_0^{t_f} \frac{\langle x^2 \rangle}{2} \frac{d\kappa}{dt}dt.
\label{eq:w1}
\end{equation}
According to our ansatz (\ref{eq:intstate}), $\langle x^2 \rangle=1/(2\alpha(t))$, and using the relation (\ref{eq:combFKsimply}), Eq.~(\ref{eq:w1}) 
can be written in the following form \cite{ScSe07,ScSe08}:
\begin{eqnarray}
\langle W \rangle \, 
& = &  \, \int_0^{t_f} \frac{1}{4\alpha} \frac{d\kappa}{dt}dt= \int_0^{t_f} \frac{\kappa}{4} \frac{\dot \alpha}{\alpha^2}dt \nonumber \\ 
& = & 
\frac{k_BT}{2}\log \left( \frac{\kappa_f}{\kappa_i}\right) \,+\, k_B T \, \frac{\tau_{\rm relax}}{t_f} \,\eta,
\label{eq:w13}
\end{eqnarray}
where $\tau_{\rm relax}=\gamma/\kappa_f$ and $\eta$ is a numerical factor given by
\begin{equation}
\eta = \frac{\alpha_f}{4} \int_0^1 \frac{1}{\alpha^3}\left(  \frac{d\alpha}{ds}\right)^2ds 
= 9 \left( \frac{\Delta \kappa}{\kappa_i}\right)^2 \int_0^1\frac{s^2(1-s)^2}{(1+(\Delta \kappa/\kappa_i)(3s^2-2s^3))^3} \,ds.
\end{equation}
  {Notice that eqs.\ref{eq:w13} coincides with expressions derived in previous works \cite{Crooks_PRL_2012,Bonanca_2014} using linear response theory. } \\
For our parameters, we find $\eta\simeq 0.106$, as indicated in the main text. Interestingly, expression (\ref{eq:w13}) gives the free energy difference value 
in the limit $t_f \gg \tau_{\rm relax}$, $0.5 k_BT \log (\kappa_f/\kappa_i)$, which appears as the minimal mean work. In the opposite limit, we have a time-energy 
relation: $t_f\langle W \rangle=k_BT \tau_{\rm relax} \eta$. We emphasize that the scaling in $1/t_f$ when $t_f \to 0$ is ansatz 
independent, while the specific value of the $\eta$ parameter depends on the ansatz. It can be shown that the lowest 
$\eta$ value for all admissible protocols, is $(\sqrt{\kappa_i/\kappa_f}-1)^2$, which gives $3/2-\sqrt{2}\simeq 0.086$ here.
Thus, our protocol, although sub-optimal in terms of mean work, nevertheless has an $\eta$ value close
to the best achievable.}

  {{\bf DAS} The data that support the plots within this paper and other findings of this study are available from the corresponding author upon request}

\newpage

\newpage

\newpage

\newpage



\section*{Supplementary material}

# Supplementary information: Engineered Swift Equilibration of a Brownian particle

Ignacio A. Martínez [1], Artyom Petrosyan [1], David Guéry-Odelin[2],
Emmanuel Trizac[3], Sergio Ciliberto[1]

1 : Laboratoire de Physique, CNRS UMR5672
Université de Lyon, École Normale Supérieure,
46 Allée d'Italie, 69364 Lyon, France.
2 : Laboratoire Collisions Agrégats Réactivité, CNRS UMR5589,
Université de Toulouse, 31062 Toulouse, France
3 : LPTMS, CNRS, Univ. Paris-Sud, Université Paris-Saclay,
91405 Orsay, France

September 6, 2016

## 1 Generalized ESE protocol

When the system is manipulated by a non harmonic potential $U(x,t)$, the position distribution is no longer Gaussian, either in equilibrium or out of equilibrium. In this case, one has to solve the following Fokker-Planck equation for the over damped situation considered here

$$\partial_t \rho(x,t) \,=\, \frac{1}{\gamma}\, \partial_x \left[\rho\, \partial_x(U)\right] \,+\, D\, \partial_{xx}^2\, \rho. \qquad (1)$$

This relation is linear in $U$, so that when the target distribution $\rho(x,t)$ has been chosen, it is possible to express the associated external potential $U$ that will guarantee the desired dynamics, as

$$U(x,t) \,=\, -k_B T \log \rho(x,t) \,+\, k_B T \int^x dy \left\{ \frac{\int^y \partial_t \rho(z,t) dz}{\rho(y,t) D} \right\}. \qquad (2)$$

As an illustration, considering $\rho$ of Gaussian form, we recover all the results derived in the main text, and in particular the dynamical equation connecting $\alpha$ and $\kappa$. Beyond the harmonic case, taking $\rho(x,t)$ of the form $\exp(-\beta x^4)$ (up to normalization), we can compute explicitly the confining potential $U/(k_B T) = \beta x^4 + A x^2$. Once the evolution law $\beta(t)$ is chosen, the only unknown $A$ follows from $A = (4D)^{-1} d\log\beta/dt$. Hence, the decompression of a state in

1

$\exp(-\beta x^4)$, which requires that $\dot\beta$ be negative in some time window, yields a negative value of $A$, which corresponds to a drive with a bistable potential $U$. The argument readily extends to target densities of the form $\rho(x,t) \propto \exp(-\beta x^n)$ where we find a driving potential $U/(k_B T) = \beta x^n + Ax^2$ with $A = (nD)^{-1} d\log\beta/dt$.

## 2 Data analysis

### 2.1 System dynamics

Figure S1 shows a set of trajectories for both processes, ESE and STEP. This highlights the impossibility to define equilibrium following solely a single trajectory. Equilibrium is indeed a statistical notion.

The linearity of Langevin equation in the case of a harmonic potential guarantees the Gaussianity of the position probability density function $\rho(x,t)$. Experimentally, the Gaussianity of a data set is quantified through its kurtosis. This parameter is defined from the centered fourth moment $\mu_x$ and the standard deviation $\sigma_x$ of the distribution: $\text{Kurt}(x) = \mu_x/\sigma_x^4$. For a Gaussian distribution, we have $\text{Kurt}(x) = 3$. The experimental values of the kurtosis along the ESE protocol are displayed in Fig. S2, illustrating how the distribution remains Gaussian during the whole process.

### 2.2 Experimental uncertainties.

The system is calibrated with standard techniques such as equipartition theorem and power spectral density [1]. The calibration factor of the photodiode is $S = (2666 \pm 3)$nm/V. The absolute errors of the system's observables are $\Delta\kappa = 0.03$pN/$\mu$m and $\Delta x = 0.1$nm. Then, to obtain the total error of the measures, the statistical uncertainty is calculated with a confidence interval of 99 % over the $N = 2 \cdot 10^4$ cycles. This yields an error of about 1% on the standard deviation of the position. Finally, in the case of the cumulative energetics of both processes, fluctuations are intrinsic as the system is in contact with a thermal bath. Thermal fluctuations produce a constant exchange of energy between the system and its environment, as heat, even with no change in the control parameter. The variance of heat is consequently larger than the variance of the work.

### 2.3 Range of validity of the method

How fast can we run the ESE protocol? In our particular experimental case, the shortest time is set by the validity of the model, together with experimental limitations. We start with the first point. Our description is overdamped, and neglects the inertial term in the Langevin equation. This is admissible provided we do not tamper the rapid ballistic regime, which requires

$t_f > m/\gamma \simeq 1\mu$s, where $m$ is the colloid mass and $\gamma$ is the viscosity term [2]. For shorter times, the underdamped extension of the problem must be taken into account.

We next address the experimental limitations of our setup. We chose the process time as 0.5 ms as a compromise between the relaxation time, the maximum acquisition frequency $f_{\text{acq}} = 20$ kHz and the maximum stiffness we can achieve, $\kappa_{\max} \simeq 50\text{pN}/\mu$m. The time evolution of the trap stiffness is indeed non-monotonous, reaching an extremum that significantly exceeds the final value. The experimental stiffness being proportional to the optical power available, a more powerful laser will allow for a decrease of the ESE time $t_f$.

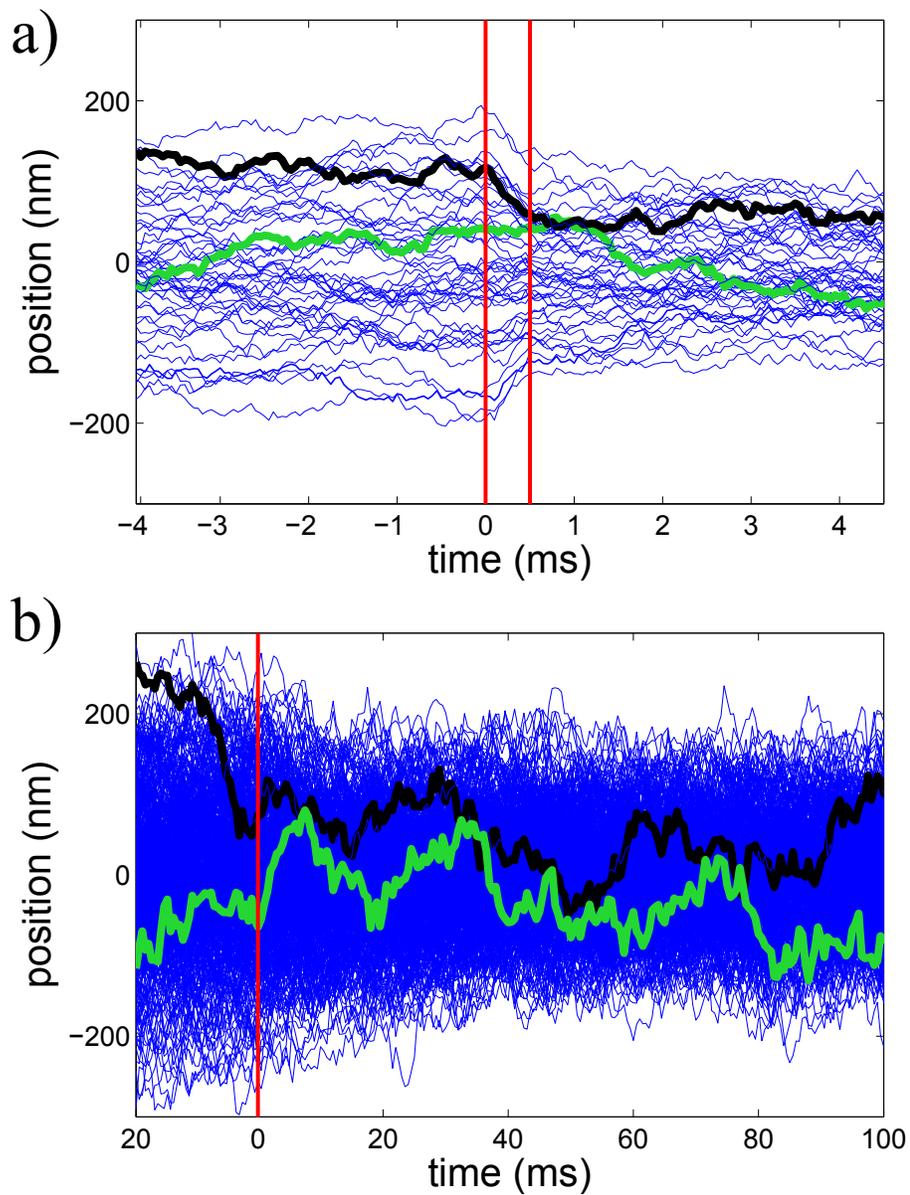

Figure S1: **Different sets of trajectories for the two different processes.** a) ESE protocol. b) STEP protocol. In each case, we highlight two particular trajectories (thick green and black lines) to show the difficulty of observing equilibration at such a level of description. In a), vertical red lines represent the initial and final times of the protocol. In b), the vertical red line represents the time instant when the potential landscape is abruptly changed.

4

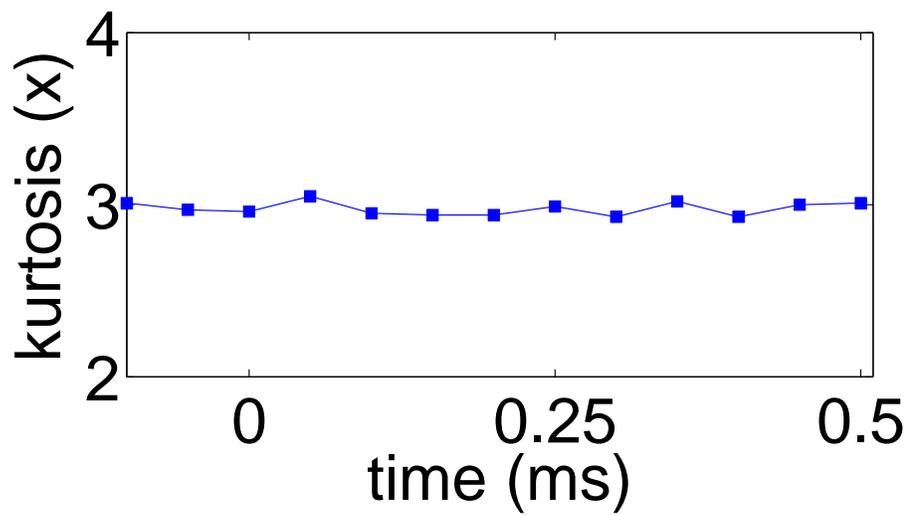

Figure S2: Experimental measure of the position distribution kurtosis Kurt($x$) during the ESE protocol. Statistical errors are below the symbol size.



\end{document}